# STUDY OF BEAM SYNCHRONIZATION AT JLEIC

*V.S. Morozov, Ya.S. Derbenev, J. Guo, A. Hutton, F. Lin, T. Michalski, P. Nadel-Turonski,*
*F. Pilat, R. Rimmer, T. Satogata, H. Wang, Y. Zhang*
Thomas Jefferson National Accelerator Facility (JLab), Newport News, VA 23606
*B. Terzic,* Old Dominion University, Norfolk, VA 23529
*U. Wienands,* SLAC National Accelerator Laboratory, Menlo Park, CA 94305

## TABLE OF CONTENTS

**Abstract**

The ion collider ring of Jefferson Lab Electron-Ion Collider (JLEIC) accommodates a wide range of ion energies, from 20 to 100 GeV for protons or from 8 to 40 GeV per nucleon for lead ions [1, 2]. In this medium energy range, ions are not fully relativistic, which means values of their relativistic beta are slightly below 1, leading to an energy dependence of revolution time of the collider ring. On the other hand, electrons with energy 3 GeV and above are already ultra-relativistic such that their speeds are effectively equal to the speed of light. The difference in speeds of colliding electrons and ions in JLEIC, when translated into a path-length difference necessary to maintain the same timing between electron and ion bunches, is quite large. In this paper, we explore schemes for synchronizing the electron and ion bunches at a collision point as the ion energy is varied.

## 1. Introduction and problem identification

Some of the beam synchronization approaches discussed in this paper including bunch number change, adjustment of the electron ring circumference and RF frequency, etc. were earlier summarized in Refs. [1, 3]. Fundamentally, to ensure collisions, the arrival times of the electron and ion bunches at the interaction point (IP) must be the same:

$$T_0 = T_{0e} = \frac{\lambda_{0e}}{c} = \frac{L_{0e}}{n_{0e}c} = T_{0i} = \frac{\lambda_{0i}}{\beta_{0i}c} = \frac{L_{0i}}{n_{0i}\beta_{0i}c}, \qquad (1)$$

where the index 0 refers to a synchronized situation, $T_{0e}$ and $T_{0i}$ are the timings between the electron and ion bunches, respectively, $\lambda_{0e}$ and $\lambda_{0i}$ are the spacings between the electron and ion bunches, $L_{0e}$ and $L_{0i}$ are the circumferences of the electron and ion rings, $n_{0e}$ and $n_{0i}$ are the harmonic numbers of the electron and ion rings, $\beta_{0i}$ is the relativistic beta of ions and $c$ is the speed of light.

While the circumferences of the JLEIC collider rings ($L_{0e}$ and $L_{0i}$) can be adjusted to provide the same timings ($T_{0e} = T_{0i} = T_0$) between the electron and ion bunches at one particular ion energy (determined by $\beta_{0i}$) – such that an electron bunch that collides with an ion bunch at the interaction point on one turn will collide with the same or another ion bunch on the next turn – this matched condition cannot be maintained for the whole ion energy range. At other energies (when $\beta_i \neq \beta_{0i}$), the colliding bunches may miss each other due to the difference in timing:

$$T_{0e} = \frac{\lambda_{0e}}{c} = \frac{L_{0e}}{n_{0e}c} \neq T_i = \frac{\lambda_{0i}}{\beta_i c} = \frac{L_{0i}}{n_{0i}\beta_i c}. \qquad (2)$$

This is the beam synchronization issue. Multiple collision points may further complicate the situation.

Table 1: Ion path-length adjustment as a function of ion momentum.

| $p$ (GeV/c/u) | $\beta_i$ | $\Delta L_i$ (m) |
|---|---|---|
| 100 | 0.999956 | 0.000 |
| 90 | 0.999946 | -0.022 |
| 80 | 0.999931 | -0.053 |
| 70 | 0.99991 | -0.099 |
| 60 | 0.999878 | -0.169 |
| 50 | 0.999824 | -0.284 |
| 40 | 0.999725 | -0.498 |
| 30 | 0.999511 | -0.958 |
| 20 | 0.998901 | -2.272 |
| 10 | 0.995627 | -9.324 |

Table 1 demonstrated the degree of the beam synchronization issue in JLEIC. Assuming an ion ring circumference of 2153.78 m, a nominal bunch number of 3422 (a bunch spacing of 62.94 cm and an RF frequency of 476.3 MHz) and the fact that the two collider rings are synchronized at a proton momentum of 100 GeV/c, Table 1 shows the ion path-length adjustment necessary to compensate the change of ion velocity as a function of proton momentum (or momentum per nucleon for heavier ions). The total path-length adjustment needed to cover the 20 to 100 GeV/c proton momentum range is about 2.27 m. For heavy ions, the necessary path-length adjustment is even larger. For example, 10 GeV/c/u lead ions need a path-length difference of about 9.32 m. Clearly, conventional path-length adjustment schemes based on a chicane or a dog-leg type magnetic system are not feasible to handle such a large path-length difference. In this paper, we discuss possible scenarios of addressing this issue and identify the most suitable one as a baseline.

## 2. Bunch number variation

## 2.1 Description

One technique for mitigating the synchronization issue is bunch number variation. Suppose that the two collider rings are synchronized for 100 GeV/c protons. Suppose also that the two rings have equal bunch numbers of 3422 at an RF frequency of 476.3 MHz. The ion collider ring circumference is 2153.78 m. Note that the circumferences of the two rings are not exactly equal. According to Eq. (1), the electron ring circumference is slightly longer:

$$L_{0e} = \frac{L_{0i}}{\beta_{0i}}. \tag{3}$$

Under this matched condition, the electron and proton revolution times are equal; therefore, in case of only one IP, an electron bunch will always collide with the same particular proton bunch at the IP while both bunches are circulating in opposite directions in their own rings.

When the proton momentum in JLEIC varies between 20 to 100 GeV/c, so does the proton velocity ($\beta_i c$) and therefore the timing between the proton bunches in the ion collider ring ($T_i$). As a consequence, the synchronization condition of Eq. (1) is no longer valid:

$$T_i = \frac{L_{0i}}{n_{0i}\beta_i c} \neq T_{0e} = \frac{L_{0e}}{n_{0e} c}. \tag{4}$$

Comparing Eqs. (1) and (4), one can see that synchronization can be restored by choosing $\beta_i$ and $n_i$ such that

$$n_i \beta_i = n_{0i} \beta_{0i}. \tag{5}$$

This means that, at certain discrete energies, the difference of revolution times happens to be exactly an integer multiple of the bunch timing (or, equivalently, the needed path length adjustment equals to an integer number of bunch spacings). Thus, the synchronization condition can be restored by storing additional bunches in the ion collider ring. In this case, an electron bunch colliding with one proton bunch at the IP, after one complete revolution, will collide with a different (*n*-th) bunch down the proton bunch train. The ion energies constituting this discrete set are called harmonic energies. The procedure of adding one or more bunches to one of the rings is sometimes also called a harmonic jump.

Table 2: Harmonic ion momenta in JLEIC with 100 GeV/c as the reference momentum.

| $n_i$ | $\beta_i$ | $p$ (GeV/c/u) |
|---|---|---|
| 3422 | 0.999956 | 100.00 |
| 3423 | 0.999664 | 36.18 |
| 3424 | 0.999372 | 26.46 |
| 3425 | 0.99908  | 21.86 |
| 3426 | 0.998788 | 19.04 |
| 3427 | 0.998497 | 17.09 |
| 3428 | 0.998206 | 15.64 |
| 3429 | 0.997915 | 14.51 |
| 3430 | 0.997624 | 13.59 |
| 3431 | 0.997333 | 12.82 |
| 3432 | 0.997042 | 12.17 |
| 3433 | 0.996752 | 11.61 |
| 3434 | 0.996462 | 11.12 |
| 3435 | 0.996172 | 10.69 |
| 3436 | 0.995882 | 10.31 |
| 3437 | 0.995592 | 9.96 |

Table 2 lists the first several harmonic momenta of ion beams in the JLEIC collider ring assuming 100 GeV/c to be the reference momentum. Depending on the synchronization scheme and technical requirements, it may be more convenient to choose a different reference momentum, e.g. 60 GeV/c. However, the total path length and RF frequency adjustments necessary to cover a certain ion momentum range are independent of the choice of the reference momentum.

It can be seen from Table 2 that the harmonic energies are distributed more densely at low ion energies, particularly below 20 GeV/u. In the ion momentum range from 20 to 36 GeV/c/u, there are still three harmonic energies. This means that there should be enough energy choices for EIC physics goals below 36 GeV/c/u. For heavy ions like lead, since their energy range in JLEIC is only up to 40 GeV/c/u, the bunch number variation scheme already provides a working solution.

Concerning the intermediate energies and, particularly, the momentum gap from 36 to 100 GeV/c/u for light ions, while the bunch number variation technique does not provide a complete solution, it greatly improves the situation by limiting the maximum necessary path length adjustment to plus or minus a half of the bunch spacing, which is not very large due to JLEIC's high repetition rate. Table 3 shows the ion path-length adjustment as a function of ion momentum in case of bunch number variation. Compared to Table 1, the issue is significantly reduced.

Some of the implications of bunch number variation are discussed in the rest of this section. Some of the possible solutions of the beam synchronization issue in JLEIC with and without bunch number variation are discussed in the rest of this report.

Table 3: Ion path-length adjustment as a function of ion momentum in case of bunch number variation assuming 100 GeV/c to be the reference momentum.

| $p$ (GeV/c/u) | $\beta_i$ | $n_i$ | $\Delta L_i$ (m) |
|---|---|---|---|
| 100 | 0.999956 | 3422 | 0.000 |
| 90 | 0.999946 | 3422 | -0.022 |
| 80 | 0.999931 | 3422 | -0.053 |
| 70 | 0.99991 | 3422 | -0.099 |
| 60 | 0.999878 | 3422 | -0.169 |
| 50 | 0.999824 | 3422 | -0.284 |
| 40 | 0.999725 | 3423 | 0.132 |

## 2.2 Improvement of detector performance

Non-pair-wise collisions resulting from bunch number variation and/or having multiple IPs, in fact, offer a number of advantages for physics measurements. In a pair-wise collider, it is important to independently keep track of the figure-of-merit (FOM) of each bunch pair $i$, $(q_e q_p P_e^2 P_p^2)_i$ as a function of time. If the bunch charges $q$ and polarizations $P$ are randomly distributed, the product of the averages is not a good estimate of the sum of the individual terms. On the other hand, with non-pair-wise collisions where every bunch in one ring collides with every other bunch in the other at a time scale of seconds, the product of the averages of each quantity (over all the bunches in each train) is exactly the FOM. From the point of view of the physics measurements, each bunch train can thus be treated as a long macro-bunch, decoupling the experimental uncertainties from the micro-structure of the accelerator, offering a way to reduce the systematic uncertainty of the measurements. This is important for an EIC, where the precision of some experiments will be limited by statistics (luminosity) and for others by systematics (in which case the luminosity becomes irrelevant). The most obvious impact of non-pair-wise collisions is on polarimetry, where removing the requirement on bunch-by-bunch measurements allows for more precise methods to be applied. And in a machine with a high repetition rate, this becomes particularly important since bunch-by-bunch measurements become more difficult. But the impact of non-pair-wise collisions goes beyond polarimetry. For instance, in spectator tagging experiments, where the goal is to achieve momentum resolutions on the scale of the Fermi momentum of the nucleons inside the nucleus, it allows a more robust unfolding of the structure function from the smearing effects caused by detector resolutions and the transverse beam momentum spread.

## 2.3 Impact on luminosity

A stored accelerator beam is usually not continuous but contains one or more gaps. In ion beams, they serve as abort gaps and means to prevent the electron cloud instability. In electron beams, they are primarily used to avoid the fast ion instability. It has also been suggested that they can be used to correct the transients in RF cavities.

In colliders with pair-wise bunch collisions, the gaps in the two counter-circulating beams are usually equal and aligned to minimize luminosity loss. Since there are no collisions during the gaps, the fractional luminosity loss in comparison to the case of fully-filled rings ($\Delta\mathcal{L}/\mathcal{L}_0$) is equal to the fractional size of each ring's gap (*g*), e.g. a 5% gap in each ring reduces the luminosity by 5%. However, in a collider with non-pair-wise bunch collisions, the gaps are constantly sliding with respect to each other and the luminosity ($\mathcal{L}$) becomes proportional to the product of the filled fractions of the orbits, i.e. $\mathcal{L} \sim (1-g)^2$, resulting in a fractional luminosity loss of

$$\frac{\Delta\mathcal{L}}{\mathcal{L}_0} = (1-g)^2 - 1 \approx -2g \ . \tag{6}$$

Thus, the luminosity loss due to gaps is almost doubled. A 5% gap in each ring results in a 10% luminosity loss in case of non-pair-wise collisions. The impact of this fact on the physics program may be non-negligible and must be accounted for.

## 2.4 Dynamical issue

There has been recent study of beam-beam effects in colliders with bunch number variation (a.k.a. harmonic jump and gear change) motivated by an interest in highly energy-asymmetric collisions in RHIC [4], extending work originally done for asymmetric B-factories in the late 1980s [5]. These works conclude that a new spectrum of resonances is created in addition to traditional beam-beam resonances from synchronized colliding beams, to all transverse nonlinear orders. The dynamics was modeled in one dimension as though the beam has a spectrum of "effective tunes", where the number of independent tunes depends on the beam-beam tune shift and the least common multiple (LCM) of the number of RF buckets in each ring. For the JLEIC case, with $n_i$ and $n_e$ being o(3400) and differing by 1, this LCM may result in a very large number of tunes, essentially making it impossible to avoid linear and nonlinear resonance conditions for all combinations.

Linear resonances that can drive coherent transverse centroid instabilities in the beams are the most concerning, as dipole and quadrupole field errors are unavoidable. These resonances occur when a linear combination of tunes is an integer multiple of 0.5. In [4], these resonances are observed even with a small number of bunches, and the conclusion is that a bunch-by-bunch transverse damper with a damping time of *o*(10 turns) is required. This result was confirmed with a separate simple rigid bunch simulation by T. Satogata [6]. Additional nonlinear resonances that distort beam moments to higher orders are also present, creating spontaneous emittance growth and luminosity reduction. In simple analysis and simulations, these also have much shorter growth timescales than characteristic damping times from electron cooling or synchrotron radiation damping.

However, to this date, other effects which may mitigate these concerns are not included in simulations of beam-beam effects with bunch number variation. In particular, chromatic tune spread and associated Landau damping may create enough mixing of beam distributions to lessen the effects of these beam-beam effects. This is of particular note in JLEIC since the ion beam is strongly longitudinally focused and thus has a high synchrotron tune and correspondingly faster phase space mixing. Additional nonlinearities and tune spread from chromatic sextupoles may also help mitigate these beam instabilities. As noted in the summary of [4], more detailed 6D beam-beam simulations with bunch number variation, chromatic effects, and magnet nonlinearities are needed to evaluate the beam dynamics and luminosity impact of bunch number variation in JLEIC.

## 2.5 Simulation

There presently seems to be no code that can accurately model the single-particle non-linear dynamics and the beam-beam effect at the same time for a large enough number of turns necessary to verify the long-term stability of a collider. Firstly, commonly used step-by-step integration of the particle motion through the ring lattice cannot provide the necessary large number of turns within a reasonable computation time. Secondly, solving a beam-beam interaction exactly using techniques such as multi-grid, conjugate gradient, or Fourier transform-based approaches, is inadequate and inefficient for simulating long-term beam dynamics in colliders due to their high computational cost. The need of simulating non-pair-wise collisions greatly complicates the problem even further. When the two rings have different bunch numbers, each bunch from the first ring will collide with a number of bunches from the second beam. In an extreme case, every bunch in one ring will collide with every bunch in the other ring. In case of JLEIC, this means that every beam turn will require computation of over 3000 beam-beam interactions.

A simplified simulation of non-pair-wise collisions has been reported in [4] but the paper also concluded that a more detailed strong-strong 6D beam-beam simulation, which is usually very time consuming, is needed. Development of a new GPU-based code, which should be capable of such a simulation, has recently been proposed relying on a matrix-based arbitrary-order symplectic particle tracking for beam transport and the Bassetti-Erskine approach to the beam-beam interaction for Gaussian beam distributions [7].

## 3. Synchronization options with bunch number variation

As discussed above, beam synchronization with bunch number variation requires beam path length adjustment by up to plus or minus a half of the bunch spacing (see Table 3). Since the bunch spacing $\lambda$ in JLEIC is 62.94 cm, the necessary path length adjustment is ±31.47 cm. Narrowing the range of ion energies reduces the range of path length adjustment.

### 3.1 Moving ion magnets

The advantage of moving ion ring magnets over moving electron ring magnets for path length adjustment is that there is no need for RF frequency adjustment for beam synchronization:

$$f_i = \frac{1}{T_i} = \frac{n_i \beta_i c}{L_i} = \frac{n_i \beta_i c}{L_{0i} + \Delta L} = \frac{n_{0i} \beta_{0i} c}{L_{0i}} = \frac{1}{T_{0i}} = f_{0i} = \frac{1}{T_{0e}} = f_{0e}. \quad (7)$$

There is also no need to synchronize injection from CEBAF. The main disadvantage is that moving super-conducting ion magnets is generally more technically difficult than moving warm electron magnets.

### 3.1.a Moving ion arcs

Moving magnets in the arcs reduces the range of magnet motion in comparison to moving magnets in a chicane but one then has to deal with a large number of magnets.

The total bending angle $\theta$ of the ion collider ring is 523.4°. Assuming uniform radial movement of the arcs, the total range of radial shift of all arc elements is

$$\Delta R = \lambda / \theta \approx 69 \text{ mm}. \quad (8)$$

Clearly, this cannot be done by simply moving the beam in the magnet apertures. With about 256 gaps between the arc dipoles and quadrupoles, the maximum required gap change is about 2.5 mm. Note that the assumption of uniform arc expansion or contraction requires a dogleg at each end of the arc with a transverse shift range of 69 mm.

Another option for moving ion arc magnets is to move a 180° arc section as a whole as illustrated in Fig. 1. The advantage is that only two straight sections at the ends of the semi-circular section are being changed while all other straight sections between the magnets remain fixed.

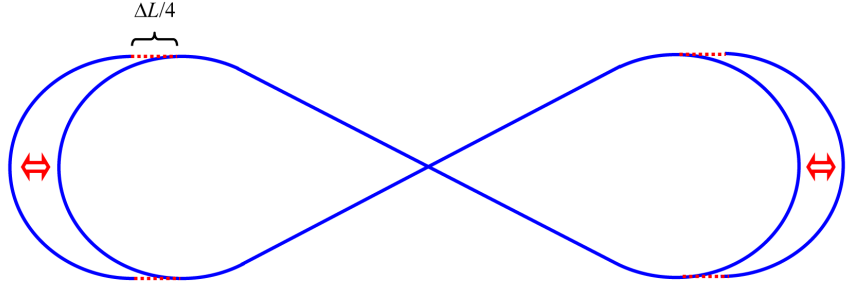

Figure 1: Moving whole 180° arc sections.

### 3.1.b Movable ion chicane (with dipoles not exceeding their nominal bending angles)

Adjusting the path length by moving magnets in a chicane requires movement of a relatively small number of magnets but the range of movement in comparison to moving the whole arcs is larger.

The chicane option described in this section does not require the chicane dipoles to exceed the nominal bending angles of the arc dipoles. Since the dipoles are rectangular, this means that there are no sagitta issues and regular arc dipoles can be used for such a chicane.

Consider an arc section consisting of a number of regular arc FODO cells with the exception that the two edge and two middle dipoles bend by angles different than regular FODO dipoles. The bending angles of the edge and middle dipoles are varied but the sum of the bending angles of one edge dipole and one middle dipole is always equal to the bending angle of a single regular FODO dipole. This way the total bending angle of the whole section remains fixed. We also require that the cord of this arc section stays fixed. Then the geometry of the rest of the ring does not change. The path length change in such a chicane comes from redistributing bending between the edge and middle dipoles. When the bend of these four special dipoles is concentrated in the edge dipoles, the chicane is in its shortest configuration. When the bend is concentrated in the middle dipoles, the path length is the longest. The geometry of a chicane consisting of three FODO cells is shown in Fig. 2. The parameters of the chicane FODO cells are the same as of the regular arc FODO cells: 22.8 m nominal FODO cell length, up to 3 T 8 m long dipoles with 3.4 m separation. Figure 2 shows that a path length adjustment of ±10 cm requires transverse magnet shift of about ±60 cm. The maximum change in magnet spacing is 20 mm. A three-FODO-cell chicane requires movement of four dipoles and five quadrupoles. Parameters of chicanes with other numbers of FODO cells are summarized in Table 4. Multiple such chicanes and/or greater transverse magnet shifts would be needed to provide the necessary ±31.47 cm path length adjustment.

Each of the chicanes in Table 4 effectively requires introduction of an extra FODO cell. It has been shown that the optics of a chicane can be matched to the regular arc FODO structure. However, due to this matching, the chicane section cannot be used efficiently for global correction of nonlinear dynamics.

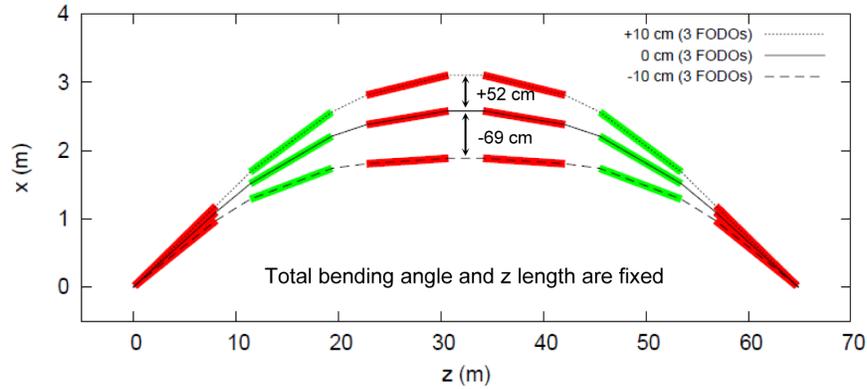

Figure 2: Three-FODO-cell ion chicane (with dipoles not exceeding their nominal bending angles).

Table 4: Parameters of ion chicanes (with dipoles not exceeding their nominal bending angles).

| # of FODOs | Length | # of magnets moved | Path length change | Radial shift range | Inter-magnet gap change |
| --- | --- | --- | --- | --- | --- |
| 3 | 68.4 m | 4 dipoles, 5 quads | ±10 cm | +52 / -69 cm | 20 mm |
| 4 | 91.2 m | 6 dipoles, 7 quads | ±10 cm | +36 / -39 cm | 14 mm |
| 5 | 114 m | 8 dipoles, 9 quads | ±10 cm | +28 / -29 cm | 11 mm |

### 3.1.c Movable ion chicane (with dipoles exceeding their nominal bending angles)

In the ion chicane option described in this section, some the chicane dipoles exceed the nominal bending angles of the arc dipoles. However, the maximum magnetic field of the chicane dipoles does not exceed that of the regular arc dipoles. The advantages of this option include the facts that such a chicane does not require any additional space in the arcs and is capable of efficiently providing a large path length adjustment. It may provide a path length change sufficient for beam synchronization without a harmonic jump. The main disadvantage is that, as the dipole bending angle exceeds the nominal one, the sagitta also exceeds the one in a regular dipole. The sagitta grows roughly linearly with the bending angle. Thus, a special dipole design and/or a larger number of shorter dipoles are needed. Such a chicane may be applicable for adjusting the electron path length as well. However, impact of the stronger bending on the synchrotron radiation power, equilibrium emittance and polarization must be considered.

Consider an arc section consisting of a number of FODO cells that has a regular lattice and all dipoles fully powered the top proton momentum of 100 GeV/c. As the proton momentum goes down to 20 GeV/c, the total field integral of the whole chicane is reduced accordingly but, instead of the usual uniform field reduction, it is first reduced in the central pair of dipoles down to zero and then moving on to the next pairs of dipoles nearest the center. The total bending angle and the cord of the chicane are fixed. The smaller bend of the central dipoles is compensated by larger bending angles of the outer dipoles. Since the momentum change in the ion ring is a factor of 5, it is natural to consider a chicane consisting of 5 FODO cell. In its longest configuration at 100 GeV/c, all of its dipoles are at the maximum field. In the shortest configuration at 20 GeV/c, the two edge dipoles remain at the maximum field while the eight central dipoles are turned off. These two extreme cases are illustrated in Fig. 3 using baseline FODO parameters described in Section 3.1.b. A transverse shift of 8.26 m along with a change of each magnet spacing by 11 cm gives a path length change of 1.97 m. Smaller path length adjustments are, of course, possible with smaller transverse magnet shifts. The optics of the chicane section needs to be adjusted for different chicane settings as for the option in Section 3.1.b.

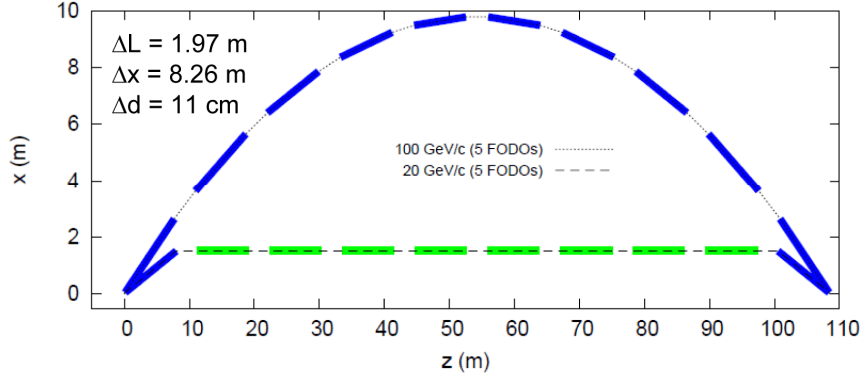

Figure 3: Five-FODO-cell ion chicane (with dipoles exceeding their nominal bending angles).

## 3.2 Moving electron magnets and tuning RF in both rings

Another option for synchronizing electron and ion bunch collisions at $\beta_i \neq \beta_{0i}$ is adjustment of the electron path length such that:

$$T_i = \frac{L_{0i}}{n_{0i}\beta_i c} = T_e = \frac{L_e}{n_{0e}c} = \frac{L_{0e} + \Delta L_e}{n_{0e}c}. \qquad (9)$$

The advantage of this approach is that moving warm electron magnets is technically easier than moving super-conducting ion ones. However, you may notice the bunch timing and therefore the bunch frequency are no longer equal to the nominal ones:

$$T_i = T_e \neq T_{0i,e}, \; f_i = f_e = \frac{1}{T_e} = \frac{n_{0e}c}{L_{0e}+\Delta L_e} \approx \frac{n_{0e}c}{L_{0e}}(1-\frac{\Delta L_e}{L_{0e}}) = f_{0e}(1-\frac{\Delta L_e}{L_{0e}}) \neq f_{0i,e}. \qquad (10)$$

Thus, the RF frequency has to be adjusted in both the electron and ion rings along with the electron path length. Employing ion harmonic jumps again limits the necessary electron path length adjustment to plus or minus a half of the bunch spacing (±31.47 cm) and hence, according to Eq. (10), the necessary frequency adjustment. Table 5 lists the required electron path-length change and RF frequency adjustment as functions of the ion momentum for the case of variable ion bunch number assuming 100 GeV/c to be the reference ion momentum. Note that a different reference momentum may be chosen for convenience. However, this does not change the total required path length and RF frequency adjustment ranges.

Table 5: Electron path-length change and adjustment of RF frequency of both collider rings as functions of the ion momentum in the case of variable ion bunch number assuming 100 GeV/c to be the reference ion momentum.

| $p$ (GeV/c/u) | $\beta_i$ | $n_i$ | $\Delta L_e$ (m) | $\Delta f$ (kHz) |
|---|---|---|---|---|
| 100 | 0.999956 | 3422 | 0.000 | 0 |
| 90 | 0.999946 | 3422 | -0.022 | -4.9 |
| 80 | 0.999931 | 3422 | -0.053 | -11.8 |
| 70 | 0.99991 | 3422 | -0.099 | -21.8 |
| 60 | 0.999878 | 3422 | -0.169 | -37.3 |
| 50 | 0.999824 | 3422 | -0.284 | -62.9 |
| 40 | 0.999725 | 3423 | 0.132 | 29.1 |

### 3.2.a Moving electron arcs

Similarly to moving ion arcs (Section 3.1.a), moving magnets in the electron arcs reduces the range of magnet motion in comparison to moving magnets in a chicane but one then has to deal with a large number of magnets.

The total bending angle $\theta$ of the electron collider ring is 523.4°. Assuming uniform radial movement of the arcs, the total range of radial shift of all arc elements is again

$$\Delta R = \lambda / \theta \approx 69 \text{ mm} . \qquad (11)$$

Clearly, this cannot be done by simply moving the beam in the magnet apertures. With about 336 gaps between the regular arc dipoles and quadrupoles, the maximum required gap change is about 1.9 mm. Note that the assumption of uniform arc expansion or contraction requires a dogleg at each end of the arc with a transverse shift range of 69 mm.

Another option for moving electron arc magnets is to move a 180° arc section as a whole (see Fig. 1 in Section 3.1.a). The advantage is that only two straight sections at the ends of the semi-circular section are being changed while all other straight sections between the magnets remain fixed.

### 3.2.b Movable electron chicane

The concept of a movable electron chicane is exactly the same as that of an ion chicane described in Section 3.1.b. A five-FODO-cell option with a nominal path length change of ±10 cm is shown in Fig. 4. The transverse magnet shift needed for a 20 cm path length adjustment is about 94 cm with the change of inter-magnet spacing of about 11 mm. A number of such chicanes or a larger transverse shift is needed to provide the total needed path-length change. As in the ion chicane case, the chicane optics has to be tuned to account for the change in the bending angles and, for this reason, is not convenient for chromaticity compensation. Additionally, for electrons, impacts on the equilibrium emittance and polarization life time should be considered although they are expected to be small.

Another option for adjusting an electron path-length chicane is to keep the magnets connected rigidly in each half of the chicane. Each half is pivoted about the respective chicane end point. The path length is adjusted by change in spacing between the two central dipoles. The advantage of this option is that there is no change in any of the other inter-magnet spacings. Therefore, no special bellows are needed there. On the other hand, the change in distance between the central magnets is relatively large and thus requires replacement of a short section of the vacuum chamber at that location.

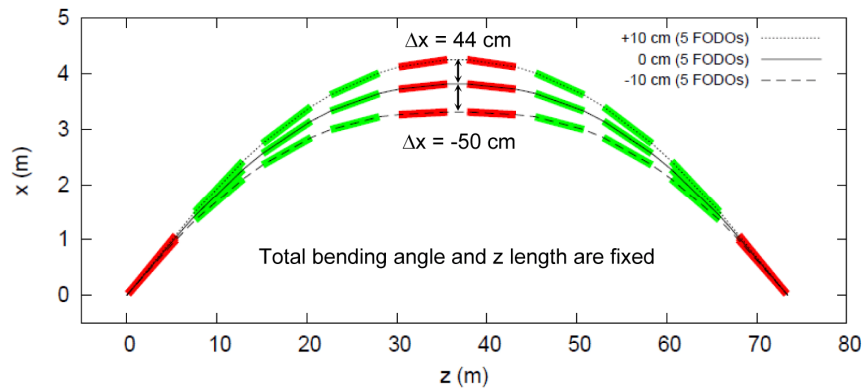

Figure 4: Five-FODO-cell movable electron chicane.

### 3.3 Bypass lines for electron path length "jumps" and RF adjustment in both rings

Suppose there are two (or more) beam lines in one section of the electron ring as illustrated in Fig. 5. At any given time, the electron beam passes through only one of them. The difference in the path lengths of the two beam lines produces a path length "jump". Several such "jumps" can be distributed around the ring. A few possible configurations of path length "jumps" are summarized in Table 6. Configuration 1 is illustrated in Fig. 6. It has one 4 cm and three 8 cm "jumps" plus two adjustable CEBAF-like chicanes. Different "jumps" can be combined with each other and with small adjustable chicanes to produce the necessary path length adjustment as illustrated in Table 7 for Configuration 2. The small chicanes are conventional chicanes, similar to those in CEBAF, with fixed magnet positions providing cm scale path length adjustments. One arc provides 32 cm path length adjustment. Two arcs provide the whole range of required path length adjustment.

Table 6: Possible configurations of path length "jumps".

| Configuration | 1 | 2 | 3 | 4 |
|---|---|---|---|---|
| | Path length adjustment (cm) | | | |
| Sum of chicanes | (±1)×2 = 4 | 4.5 | 5 | 6 |
| Jump 1 | 4 | 4.5 | 5 | 6 |
| Jump 2 | 8 | 9 | 10 | 10 |
| Jump 3 | 8 | 13.5 | 12 | 10 |
| Jump 4 | 8 | | | |
| Total | 32 | 31.5 | 31.5 | 32 |

Table 7: Combining path length "jumps" and small chicanes of Configuration 2 in Table 6.

| Building blocks | Path length (cm) |
|---|---|
| Chicanes | 0 to 4.5 |
| Chicanes + Jump 1 | 4.5 to 9 |
| Chicanes + Jump 2 | 9 to 13.5 |
| Chicanes + Jump 3 | 18 to 22.5 |
| Chicanes + Jumps 2 &3 | 22.5 to 27 |
| Chicanes + Jumps 1, 2, & 3 | 27 to 31.5 |

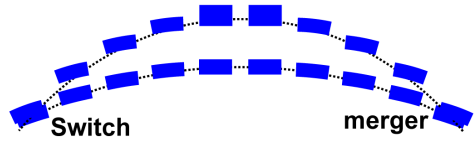

Figure 5: Two-beam-line path length "jump".

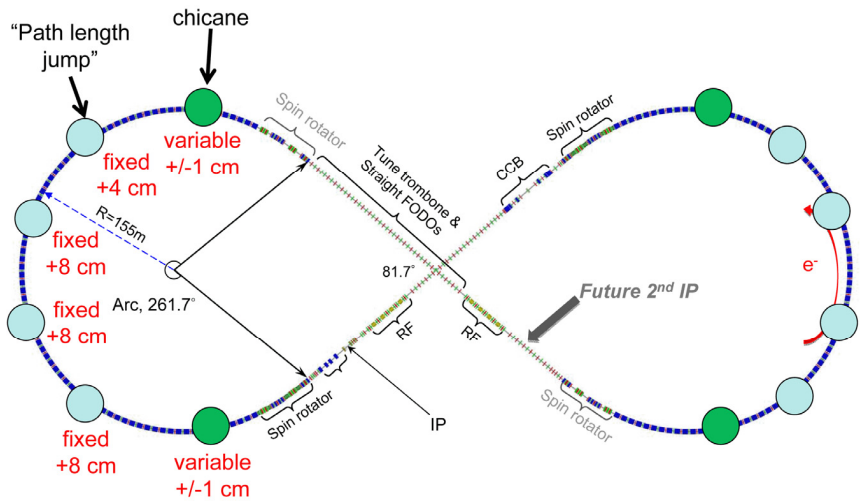

Figure 6: Path length "jumps" and small chicanes of Configuration 1 in Table 6 placed around the ring.

## 3.4 Scanning synchronization

Scanning synchronization [8] is a novel synchronization option that does not require change of the orbit length and therefore does not involve magnet movement. It can provide bunch collisions when the RF frequencies of the two rings are not equal. The position of the crossing point is adjusted for every bunch pair in such a way that it compensated the delay of arrival of one of the colliding bunches at the nominal crossing point. Since the delay time changes for every bunch pair in a periodic manner, the collision point moves periodically in time (scans) over a small longitudinal range equal to $\pm\lambda/4 = \pm15.7$ cm. The transverse offset of the crossing point is therefore $h = \pm(\lambda/4)\theta_{cr} = \pm 8$ mm. The collision point is moved by shifting one or both orbits in the detector region transversely as shown in Fig. 7. The crossing point starts at one of the longitudinal range, moves linearly to the other end and then jumps back to the original location. This periodic movement is accomplished using a pair of fast kickers. These fast kickers can be implemented as dipole-mode RF cavities fed by magnetron power sources, allowing broad-band phase and amplitude control. RF energy of the scanner can be recycled using a fast switch. Thus, scanning synchronization requires kicker technology as well as a study of its compatibility with beam dynamics and detection scheme.

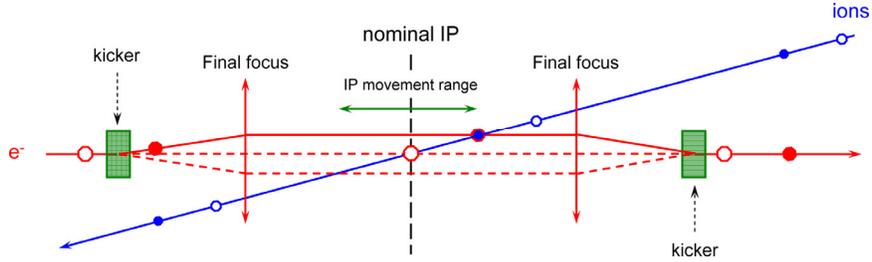

Figure 7: Schematic of scanning synchronization.

## 4. Synchronization options with the same numbers of bunches

As it will be discussed below in this section, synchronization with pair-wise bunch collisions requires much greater beam path-length adjustment and/or frequency tuning range than the case of non-pair-wise bunch collisions discussed in Section 3. The main reason for considering this option is the possibility of a dynamic instability caused by non-pair-wise collisions described in Section 2.4. This instability does not appear to be critical in the JLEIC case but is still being studied. Also, with the same numbers of electron and ion bunches, the beam gaps are synchronized minimizing luminosity loss (see Section 2.3).

### 4.1 Moving ion arcs

The same arguments apply to moving the ions arcs in the pair-wise case as in the non-pair-wise one (see Section 3.1.a). There is no need for RF frequency adjustment in either ring but one has to deal with a large number of super-conducting magnets. However, the difference from the non-pair-wise case is that the required path-length adjustment is much greater as shown in Table 1. According to Eq. (8), the required radial magnet shifts would have to be about 25 cm at 20 GeV/c/u and about 1 m at 10 GeV/c/u to provide the necessary path-length changes of about 2.3 and 9.3 m, respectively. With about 256 gaps between arc dipoles and quadrupoles, the changes in the inter-magnet spacings would have to be about 9 and 36 mm, respectively. This option is certainly not very attractive and is described here for completeness.

### 4.2 Moving electron arcs and tuning RF in both rings

When moving magnets in the electron arcs while maintaining pair-wise collisions, the required magnet movement is the same as for the ion magnets in the same regime described in Section 4.1. In addition, as discussed in Section 3.2, the frequency of RF cavities has to be adjusted in both rings to match the electron path length change. Table 8 lists the required electron path-length change and RF frequency adjustment as functions of the ion momentum. The possibility of such a large frequency change would have to be investigated. This option is also described here for completeness.

Table 8: Electron path-length change and RF frequency adjustment in both collider rings as functions of the ion momentum in the case of pair-wise collisions.

| $p$ (GeV/c/u) | $\beta_i$ | $n_i$ | $\Delta L_e$ (m) | $\Delta f$ (MHz) |
|---|---|---|---|---|
| 100 | 0.999956 | 3422 | 0 | 0 |
| 90 | 0.999946 | 3422 | -0.02 | -0.005 |
| 80 | 0.999931 | 3422 | -0.05 | -0.012 |
| 70 | 0.99991 | 3422 | -0.10 | -0.022 |
| 60 | 0.999878 | 3422 | -0.17 | -0.037 |
| 50 | 0.999824 | 3422 | -0.28 | -0.063 |
| 40 | 0.999725 | 3423 | -0.50 | -0.11 |
| 30 | 0.999511 | 3424 | -0.96 | -0.21 |
| 20 | 0.998901 | 3426 | -2.27 | -0.50 |
| 10 | 0.995627 | 3437 | -9.32 | -2.06 |

### 4.3 System of electron path-length "jumps" with simultaneous change of electron and ion harmonic numbers and RF adjustment

In this scheme, the ion bunch number changes with energy as shown in Table 9. This is done to minimize the required RF frequency change. Then, to keep the collisions pair-wise, the electron bunch number is also changed to match the ion harmonic number. This avoids the potential dynamic problem with non-pair-wise collisions. In addition, the gaps are also synchronized. The electron path length is changed to accommodate the additional bunches. Large path-length "jumps" are added to absorb the large path length change. Energies between the harmonic ones are covered by one of the schemes described in Chapter 3. Covering the ion energy range from 18 to 100 GeV requires addition of 4 electron bunches and therefore a total path length change of $4\lambda = 252$ cm. Table 10 lists possible configurations that combine large "jumps" and the scheme described in Section 3.3 to reach the necessary path-length adjustment.

Table 9: Change of the ion bunch number with energy.

| Ion energy (GeV/u) | Bunch number |
|---|---|
| 47.25 | 3416 |
| 29.28 | 3417 |
| 22.91 | 3418 |
| 19.38 | 3419 |

Table 10: Possible combination of large path length "jumps" and the scheme described in Section 3.3.

| Configuration | 1 | 2 | 3 | 4 |
|---|---|---|---|---|
| | Path length adjustment (cm) | | | |
| Sum of chicanes | 31.5 | 4 | 4.5 | 6 |
| Jump 1 | | 4 | 4.5 | 6 |
| Jump 2 | | 8 | 9 | 10 |
| Jump 3 | | 8 | 13.5 | 10 |
| Jump 4 | | 8 | | |
| Large jump 1 | 31.5 | 31.5 | 31.5 | 31.5 |
| Large jump 2 | 63 | 63 | 63 | 63 |
| Total for two arcs | 242 | 242 | 242 | 242 |

### 4.4 System of electron bypass lines with simultaneous change of electron and ion harmonic numbers and RF adjustment

The large electron path length adjustment necessary for synchronization with pair-wise collisions can be obtained using a system of electron bypass beam lines shown in Fig. 8. The edge dipoles shown in red send the electron beam down one of the 5 paths. The parameters of each path line are given in Table 11. Note that synchrotron radiation power limits and required apertures must be checked for the dipoles exceeding the nominal bending angles.

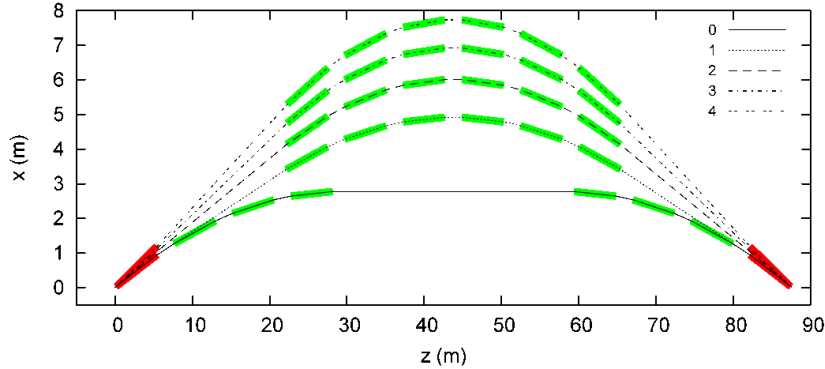

Figure 8: System of electron bypass beam lines.

Table 11: Parameters of the individual beam lines in Fig. 8.

| Line # | B1 angle (°) | B2-4 angle (°) | Long straight (m) | $\Delta x_{max}$ (m) | $\Delta L$ (m) |
|---|---|---|---|---|---|
| 0 | 2.8 | 2.8 | 31.24 | 0 | 0 |
| 1 | 2.8 | 2.8 | 2×16.88 | 2.14 | 0.315 |
| 2 | 0.74 | 3.49 | 2×17.04 | 3.23 | 0.74 |
| 3 | -0.97 | 4.06 | 2×17.19 | 4.15 | -0.97 |
| 4 | -2.47 | 4.56 | 2×17.35 | 4.96 | -2.47 |

### 4.5 Scanning synchronization

The scheme described in Section 3.4 is also applicable in the case of pair-wise bunch collisions.

## 5. Synchronization of a second interaction point

Suppose the second IP is placed symmetrically to the first one, i.e. the distance between the IPs is a half of the ring circumference. Then, if the path length is adjusted symmetrically in the two arcs and if the difference in the numbers of electron and ion bunches is an even number, synchronization of one IP using one of the techniques described above automatically synchronizes the second IP. If the bunch number difference is an odd number then the bunches miss each other at the second IP by a half of the bunch spacing $\pm \lambda / 2$. One can restore collisions by shifting the collision point to be half-way between the bunches, i.e. by $\pm \lambda / 4$. This can be done by a small adjustment of the crossing angle accompanied by an appropriate adjustment of the focal distance. Moreover, this IP shift can be distributed between the two IPs reducing the movement of an individual IP to $\pm \lambda / 8$. If it is expected that the bunch number difference will always be odd the second IP can be designed already shifted.

Another option for synchronizing the second IP with an odd bunch number difference is to adjust the path lengths in the two arcs asymmetrically. If the difference in the path lengths in the two arcs is a half of the bunch spacing $\lambda / 2$, i.e. $\pm \lambda / 4$, the second IP is also synchronized.

## 6. RF cavity tuning

### 6.1 PEP-II cavity tuning

PEP-II cavities have one fixed and one movable tuner. Each tuner has ~±500 kHz range. This is sufficient for tuning to 476.3 MHz as shown in Fig. 9. The tuning range is also wide enough for synchronization even without harmonic jump. As shown in Table 8, the frequency range needed for synchronization of 20 to 100 GeV/u ions is about ±250 kHz.

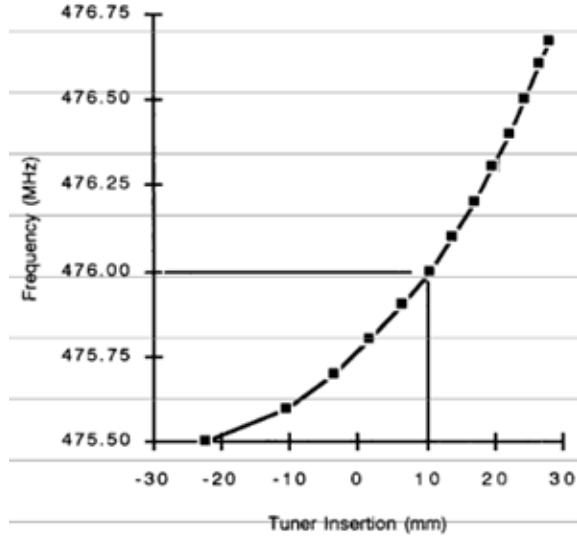

Figure 9: PEP-II cavity frequency versus tuner insertion.

## 6.2 SRF cavity tuning

The ion collider ring uses 952.6 MHz SRF cavities foreseeing a future frequency upgrade. Initially every other bucket is filled. At this frequency, changing the bunch number by 1 (changing the harmonic number by 2) requires a frequency tuning range of about ±140 kHz. The frequency range needed to cover 20-100 GeV/u ion energies without harmonic jump is about ±500 kHz. Scaling present JLab experience to 952.6 MHz gives a tuning range of about ±250 kHz. Thus, it is sufficient for synchronization with harmonic jump. Synchronizing without harmonic jump will require further R&D to allow a larger tuning range.

## 6.3 Crab cavity tuning

Extrapolating JLab SRF experience to crab cavities leads to similar conclusions as in Section 6.2.

## 6.4 Synchronization with CEBAF

Any beam synchronization scenario in the JLEIC that involves a change of the electron ring path length results in an adjustment of the RF frequency in both electron and ion collider rings. Hence, CEBAF electron injection frequency needs to be tuned to match the electron ring frequency and an appropriate path length change in the CEBAF is required accordingly. However, the current chicanes in the CEBAF can only provide a ±1 cm path length change, which is far below the requirement of ±10 cm in each arc of CEBAF. There is no doubt that one can synchronize CEBAF and electron ring RF frequencies by significant adjustments in both optics and instruments, but it may not be practical in reality. Instead, we can keep the CEBAF frequency same as that at one optimum synchronization setting and inject electron bunch trains at the off-sets of the RF phase in the electron ring at other energies provided the electron ring has a large enough longitudinal dynamic aperture. Synchrotron radiation in the electron ring will eventually damp the longitudinal phase-space distribution to the stable region around the synchrotron phase.

In all aforementioned synchronization options, the maximum RF frequency change in the electron ring is ±0.015%. Considering the injection of electron bunch train to the half electron ring each time, the maximum phase difference between the head (or tail) and the center of the bunch train is

$$\Delta\phi_{max} = 2\pi \cdot \frac{\Delta f}{f} \cdot \frac{h}{4} = \pm 45°. \qquad (12)$$

Here $\Delta f / f = \pm 0.015\%$, $h = 3416$ is the harmonic number. Since the maximum energy spread form the CEBAF is about ±0.2%, it requires that the longitudinal dynamic aperture in the electron collider ring must tolerate particles with $(\phi, \delta) = (\pm \pi/4, \pm 0.2\%)$. Figures 10 and 11 show the separatrix orbits ("fish" shape) with synchrotron phases $\phi_s$ in the JLEIC baseline design at 5 and 10 GeV, respectively. The phase space area enclosed by the separatrix orbit is the bucked area. As one can see, the bucket area is much larger than the required longitudinal dynamic aperture of $(\phi, \delta) = (\pm \pi/4, \pm 0.2\%)$ shown as the red rectangular box in each plot. In principle, all particles within this dynamic range should be stable and will be damped eventually. This will be confirmed by further simulations in the future.

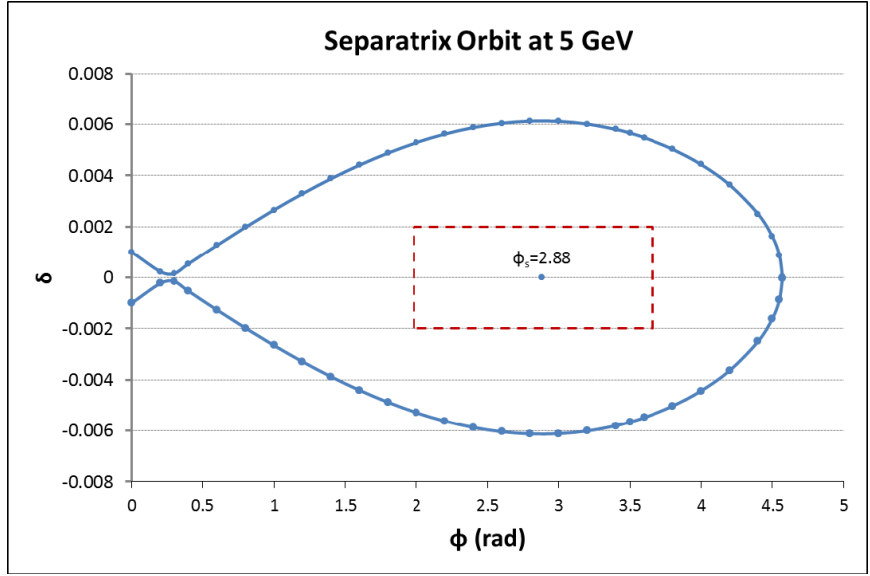

Figure 10: Separatrix orbit ("fish" shape) at 5 GeV JLEIC electron ring with the synchrotron phase $\phi_s$. The red rectangular box shows the required maxium longitudinal dynamic range for the injected electron bunch trains.

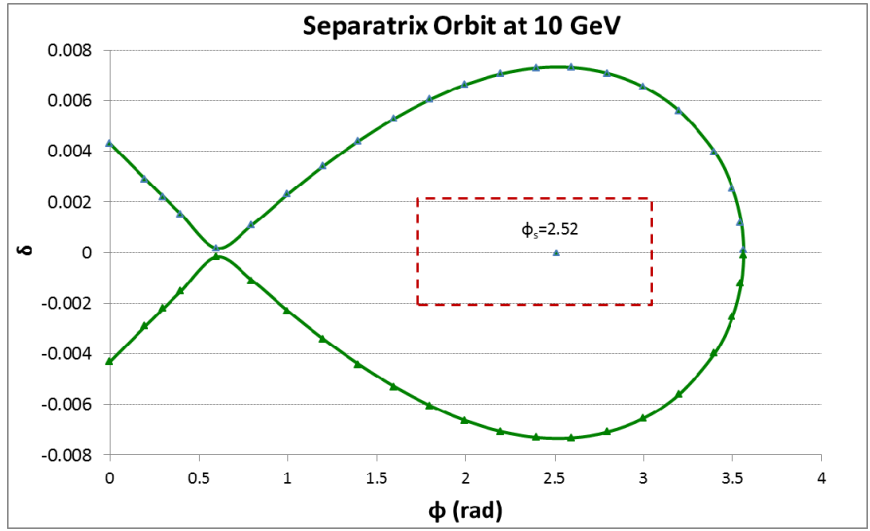

Figure 11: Separatrix orbit ("fish" shape) at 10 GeV JLEIC electron ring with the synchrotron phase $\phi_s$. The red rectangular box shows the required maxium longitudinal dynamic range for the injected electron bunch trains.

## 7. Engineering aspects of reposition magnets in collider rings

As mentioned in Sections 3 and 4, the ability to achieve synchronization is predicated on adjusting the path length in either the electron collider ring or the ion collider ring. Both rings contain large, heavy magnets in their arcs. However, the ion collider ring consists of a series of linked superconducting magnets with multiple electrical and cryogenic interfaces as well as having beam pipe and insulating vacuum. Due to the quantity of interfaces between adjacent magnet cryostats and the risk of straining the superconductor, it is not advisable to reposition these magnets. Therefore, the remainder of this section will focus on repositioning electron collider ring magnets.

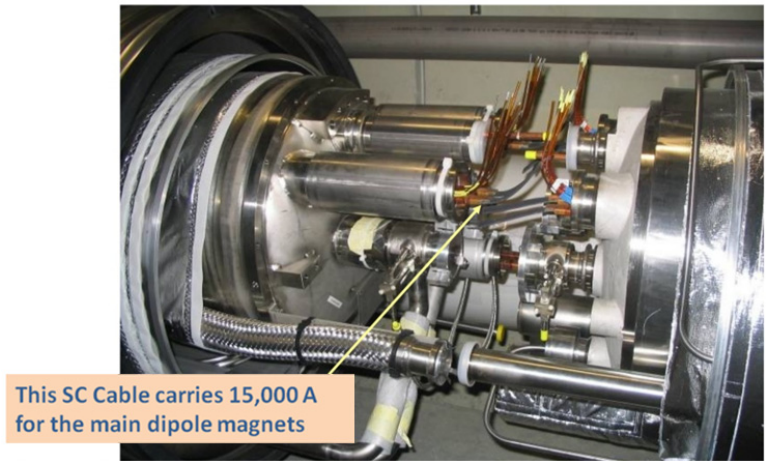

Figure 12: An example of a superconducting magnet junction.

The electron collider ring arcs consist of repurposed PEP-II high energy ring (HER) magnets, see Fig. 13. As received from SLAC, the layout consists of quadrupole/sextupole/corrector rafts which are directly coupled to one adjacent, downstream dipole magnet via a bolted flange. It is ideal to maintain this coupling, but not required. A shielded bellows is used on the upstream end of the raft to interface to the other adjacent dipole. The weight of a 5.4 m HER dipole is 16,000# and the weight of a quadrupole raft with magnets is 7,000#. In discussions with staff at SLAC, the shielded bellows between raft and dipole accommodates limited longitudinal motion (approximately 10-12 mm) and very limited transverse and vertical motion (1-2 mm). They should not be considered a dynamic element. This means each bellows should be disconnected prior to moving beamline elements in order to prevent damage. The result is a break in vacuum at all locations and considerations must be given to maintaining cleanliness while disconnected.

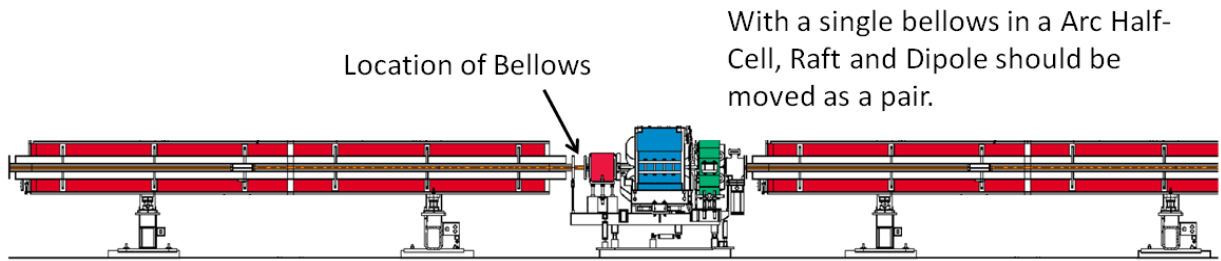

Figure 13: PEP-II arc magnets.

Two main options for repositioning electron collider ring magnets are being considered and described here, each with two sub-options.
- Per section 3.2a, Moving Electron Collider Ring Arcs;
  o Option 3.2a.1: reposition all magnets in the arc radially outward;
  o Option 3.2a.2: reposition 180° of the arc as a complete group;
- Per section 3.2b, Movable Electron Collider Ring Chicanes – 5 arc FODO cells per chicane;
  o Option 3.2b.1: reposition all magnets in 5 arc FODO cells, requiring both translation and rotation of each magnet;
  o Option 3.2b.2: pivot each half of the 5 arc FODO cells about their associated end and install a longer vacuum tube section.

Option 3.2a.1, repositioning all magnets in the electron collider arc radially outward, would require moving each raft and dipole in the arc by a fixed amount. Each arc contains 168 elements (rafts and dipoles) which must be repositioned 69 mm radially outward. The associated gap between adjacent elements is 1.9 mm. It is recommended that these individual gaps be combined at the location of the bellows, thereby halving the number of gaps but doubling the gap size to 3.8 mm. This also requires moving each arc half-cell as one unit (raft and dipole), so as to not strain the bolted flange interface. The alternative is to unbolt the raft from the dipole, move each element, then reconnect them. The specifics of this option are:

- 336 elements require repositioning. An alignment crew can reposition on average 3-6 elements per shift. This would require 56-112 crew shifts to reposition elements and additional time to break and remake the vacuum. Multiple crews per shift are used to reduce the overall duration of the activity.
- The radial repositioning of each element exceeds the adjustment of the existing mounts. A new methodology is required and would have to be duplicated for each of the 336 elements. The limited frequency of making such adjustments allows for a manual adjustment scheme using threaded adjusting rods and a low friction plane on which the elements move. Initial installation provides for a flat plane on which to reposition the elements, therefore, vertical adjustment is minimal and will be made with existing alignment adjusters.
- Connecting each arc half-cell (raft and dipole) to move as a single unit reduces the quantity of items to be repositioned for each energy change. This will maintain alignment between the raft and dipole.
- The SLAC designed bellows can accommodate the 3.8 mm gaps generated by this option.

Option 3.2a.2, repositioning 180° of the arc as a complete group, offers a significant challenge. The radius of the arc is 155.4 m. Moving a 977 m long string of elements that span the arc will not have the required rigidity to maintain alignment between all elements without having to reposition the entire tunnel along with it. The weight of each arc's entire string of elements is nearly 2 million pounds. With these considerations, moving 180° of the arc as a complete group does not appear to be a reasonable viable option.

Option 3.2b.1, reposition all magnets in each 5 arc-cell chicane, requiring both translation and rotation of each magnet. Each chicane contains 20 elements (rafts and dipoles) which must be repositioned up to a maximum of 94 cm. The associated gap between adjacent elements is about 11 mm. It is recommended that these individual gaps be combined at the location of the bellows, thereby halving the number of gaps but doubling the gap size to 22 mm. This also requires moving each arc half-cell as one unit (raft and dipole), so as to not strain the bolted flange interface. The specifics of this option are:

- 10 arc half-cells (combined raft and dipole) per chicane require repositioning. An alignment crew can reposition on average 3-6 elements per shift. This would require 2-3 crew shifts to reposition elements and additional time to break and remake the vacuum.
- The radial movement of each element exceeds the adjustment of the existing mounts. A new methodology is required and would be duplicated for each of the 10 arc half-cells per chicane. The limited frequency of making such adjustments allows for a manual adjustment scheme. Initial installation provides for a flat plane to reposition the elements, therefore, vertical adjustment is minimal and will be made with existing alignment adjusters.
- Connecting each arc half-cell to move as a single unit reduces the quantity of items to be repositioned for each energy change. This will maintain alignment between the raft and dipole.
- The 22 mm gap will not be accommodated by the SLAC designed shielded bellows. To cover the range of created gaps, 3 sets of replacement bellows will be required.
- The tunnel will require additional width in the chicane region to accommodate the lateral motion of the magnets.
- The SLAC design of the HER beamline does not accommodate vacuum valves within the arc cell. Maintaining good cleanliness practices during the magnet repositioning process is required. Experience at SLAC has proven no degradation in the achievable vacuum when these practices are maintained.

Option 3.2b.2, pivot each half of the 5 arc-cells about their associated end and install a longer vacuum tube section. Concentrating the gap at the center of the chicane minimizes the breaks in the beamline. Each half of the chicane pivots 2.8 degrees. The associated gap between the ends of the two halves of the chicane is much larger than can be accommodated by the shielded bellows, requiring an added section of vacuum chamber. The specifics of this option are:

- Each half of the chicane contains approximately 115,000# of magnets. In order to pivot the length and weight of half the chicane requires a substantial support structure. To maintain the beam centerline, this structure is recessed into the floor of the tunnel.
- Essentially two elements must be repositioned for each energy change. This would require 2-3 crew shifts to reposition elements and additional time to break and remake the vacuum.

- Concentrating the gap at a single location requires a custom section of vacuum chamber. The variation in the gap size for the range of energies will require a range of unique vacuum chambers in conjunction with the shielded bellows.
- The tunnel will require additional width in the chicane region to accommodate the lateral motion of the magnets.
- Due to recessing the movable support structure into the floor of the tunnel, a grating or suspended floor system will be required to fill the gaps around the chicane for safety reasons.
- The SLAC design of the HER beamline does not accommodate vacuum valves within the arc cell. Maintaining good cleanliness practices during the magnet repositioning process is required. Experience at SLAC has proven no degradation in the achievable vacuum when these practices are maintained.

It seems clear that the most viable options for moving magnets should focus on the warm, electron collider ring arc magnets and use chicanes. This reduces the quantity of magnets which must be repositioned and will have a smaller impact on cost to implement due to the reduced quantity of moving/positioning mechanisms. Moving arc half-cells will require 3 sets of shielded bellows to compensate for the gaps created over the range of energies. Pivoting each half of each chicane will require an investment in a robust support structure and pivot point as well as having a greater impact on the design of the tunnel in the affected regions. Either approach requires increasing the tunnel width by an additional meter.

## 8. Impact of future frequency upgrade

Doubling of the collider frequency, from 476.3 to 952.6 MHz, is foreseen in a future JLEIC luminosity upgrade. All relative frequency changes normalized to 476.3 MHz will be reduced by a factor of 2. The absolute frequency tuning ranges will remain the same. The bunch spacing $\lambda$ will be reduced by a factor of 2. Since all required path-length adjustments scale with $\lambda$, they will also be reduced by a factor of 2 making synchronization much simpler.

## 9. Conclusions and selection of a baseline scheme

As discussed above, there is a number of solutions for synchronizing bunch collisions in JLEIC when ion energy changes. In addition, some of the options may be combined if necessary. Moreover, since change in the ion beam energy is expected to happen on a half a year to a year time scale, the necessary adjustments can be made in a few days during a shutdown. There are options for both equal and different numbers of bunches in the electron and ion rings. The choice is determined by technical convenience. Since running with different bunch numbers simplifies synchronization and provides a significant physics benefit, we focus on the corresponding options. Among them, we choose the option described in Section 3.2.b, namely, movable electron chicane (involving bunch number variation, modest RF frequency adjustment, and two or three chicanes similar to that illustrated in Fig. 4), as our baseline because (a) moving warm electron magnets is technically easier than moving super-conducting ion magnets, (b) since the required movement is large enough that it requires special design even when all arc magnets are shifted, it is preferable to deal with a small number of specially designed magnets and movers, and (c) the available RF tuning range is consistent with the frequency adjustment requirements.

## Acknowledgements

This paper summarizes the synchronization schemes discussed at a series of special meetings at Jefferson Lab during July-August 2015. We are grateful to all participants of these meeting for useful discussions.

This paper is authored by Jefferson Science Associates, LLC under U.S. DOE Contracts No. DE-AC05-06OR23177 and DE-AC02-06CH11357. The U.S. Government retains a non-exclusive, paid-up, irrevocable, world-wide license to publish or reproduce this manuscript for U.S. Government purposes.